\documentclass{ws-ijmpd}
\usepackage[super,compress]{cite}
\usepackage[breaklinks]{hyperref}
\usepackage{breakurl}
\begin{document}

\thispagestyle{plain}

\def\bib{B\kern-.05em{I}\kern-.025em{B}\kern-.08em}
\def\btex{B\kern-.05em{I}\kern-.025em{B}\kern-.08em\TeX}

\title{STABILIZING SPHERICAL ENERGY SHELLS WITH ANGULAR MOMENTUM IN GRAVITATIONAL BACKGROUNDS}
\markboth{Stabilizing Spherical Energy Shells with Angular Momentum in Gravitational Backgrounds}{Stabilizing Spherical Energy Shells with Angular Momentum in Gravitational Backgrounds}

\author{I.ANTONIOU\footnote{i.antoniou@uoi.gr}}

\address{Department of Physics, University of Ioannina,GR-45110, Ioannina,Greece}

\author{D. KAZANAS\footnote{demos.kazanas@nasa.gov}}

\address{Astrophysics Science Division, NASA/Goddard Space Flight Center, Greenbelt,Maryland 20771,USA}

\author{D. PAPADOPOULOS\footnote{papadop@astro.auth.gr}}

\address{Department of Physics, Aristotle University of Thessaloniki, Section of Astrophysics, Astronomy and Mechanics,
54124 Thessaloniki, Greece}

\author{L.PERIVOLAROPOULOS\footnote{leandros@uoi.gr}}

\address{Department of Physics, University of Ioannina,GR-45110, Ioannina,Greece
}

\maketitle

%

\newcommand{\be}{\begin{equation}}
\newcommand{\ee}{\end{equation}}
\newcommand{\bea}{\begin{eqnarray}}
\newcommand{\eea}{\end{eqnarray}}
\newcommand{\al}{\alpha}
\newcommand{\bt}{\beta}
\newcommand{\gm}{\gamma}
\newcommand{\Gm}{\Gamma}
\newcommand{\dl}{\delta}
\newcommand{\Dl}{\Delta}
\newcommand{\eps}{\epsilon}
\newcommand{\zt}{\zeta}
\newcommand{\et}{\eta}
\newcommand{\thv}{\vartheta}
\newcommand{\kp}{\kappa}
\newcommand{\lm}{\lambda}
\newcommand{\ks}{\xi}
\newcommand{\rh}{\rho}
\newcommand{\rhv}{\varrho}
\newcommand{\sg}{\sigma}
\newcommand{\Sg}{\Sigma}
\newcommand{\ta}{\tau}
\newcommand{\ups}{\upsilon}
\newcommand{\ph}{\phi}
\newcommand{\Ph}{\Phi}
\newcommand{\phv}{\varphi}
\newcommand{\ch}{\chi}
\newcommand{\ps}{\psi}
\newcommand{\om}{\omega}
\newcommand{\Om}{\Omega}
\newcommand{\fdot}{\mbox{\boldmath $\cdot$}}
\newcommand{\rarrow}{\rightarrow}
\newcommand{\Rarrow}{\Rightarrow}
\newcommand{\nn}{\nonumber}
\newcommand{\varep}{\varepsilon}
%
%
%
%


\begin{abstract}
Spherical energy shells in General Relativity tend to collapse due to gravitational effects and/or due to tension effects. Shell stabilization may be achieved by modifying the gravitational properties of the background spacetime. Thus, gravastars consist of stiff matter shells with an interior deSitter space and an exterior Schwarzshild spacetime whose attractive gravity balances the interior repulsive gravity of the interior deSitter spacetime leading to a stable stiff matter shell. Similar stabilization effects may be achieved by considering rotating shells. Here we study the stability of slowly rotating fluid shells. We show that the angular velocity of the shell has stabilizing properties analogous to the repulsive deSitter gravity of the interior of a gravastar. We thus use the Israel junction conditions\cite{Israel:1966rt,Deruelle:2007pt} and the fluid equation of state of the rotating shell to construct the dynamical equations that determine the evolution of the rotating shell radius. These dynamical equations depend on the parameters of the background spacetime and on the angular velocity of the shell. Assuming a rotating interior and a Schwarzschild exterior spacetime we show that the angular velocity of the shell has interesting stabilizing properties on the evolution of its radius R. Thus rotating matter (or vacuum) shells can imitate black holes while avoiding the presence of a singularity and without the presence of an interior deSitter space.
\end{abstract}
\maketitle
\section{Introduction.}
The interaction between gravity and matter is described by the General Relativistic (GR) inhomogeneous equations. They constitute a highly non-linear system which is tractable by analytic means in a few special, usually highly symmetric, situations. A much studied geometry has been the spherical one, but even in that case the equations are in general too complex to be solved completely when there is matter present. In Ref. ~\cite{Israel:1966rt} an elegant geometric analysis is used to derive  covariant equations of motion for situations where the matter distribution is restricted to a thin surface layer with finite density per unit area. It was found that, when such a layer is spherically symmetric (shell) and expands radially, the equations simplify and reduce to simple differential equation which can be fully integrated~\cite{Hoye:1983dq}. This constitutes one of the few cases in GR that the interaction between matter and gravity can be solved exactly, even though the analysis is valid for a shell with almost zero thickness. In the approach of Ref. ~\cite{Israel:1966rt} the surface layer is described by a surface energy momentum tensor, introduced as an independent object. It can not be generalized to include finite thickness corrections, and its internal structure and dynamics are ignored. It should be pointed out however that in Ref.~\cite{Hoye:1983dq} the authors discussed the evolution of thin spherical shells\cite{Lobo:2005zu,Brady:1991np} in the context of GR starting directly from the non-vacuum Einstein field equations coupled to an arbitrary spherical matter distribution. They showed that the resulting dynamics can be solved exactly in the limit when this distribution approaches an infinitesimal thin shell reproducing and generalizing the results of Israel ~\cite{Israel:1966rt,Davis:2002gn,Battye:2001pb}. 

Rotating bodies in GR are of particular interest since they are closer to describing astrophysical objects which in general rotate, while they include additional physical considerations like dragging effects and inertial frames~\cite{Heller:1975a,DeLaCruz:1968zz}.

The effect of a rotating hollow body in GR, in analogy to Newton's famous experiment, was studied by many authors \cite{Thirring:1918uj,Lense:1918zz}. In Ref.\cite{Thirring:1918uj} Thirring considered a thin spherical shell of radius $r_0$ rotating with constant angular velocity $\omega$ with respect to an asymptotic Minkowski space. In a follow-up paper~\cite{Mashhoon:1984fj} it was pointed out that the local inertial frame is dragged along by the rotating shell under the assumption of slow rotation and small shell mass. The so-called  Lense-Thirring metric continues to be of interest~\cite{Baines:2020unr} because it may be obtained from the Kerr metric by appropriate approximations ~\cite{Adler:1975a}. Its form, in spherical coordinates reads
\begin{equation}\label{z1t}
ds^2=-(1-\frac{2m}{r})dt^2+\frac{dr^2}{(1-\frac{2m}{r})}-\frac{4J\sin^2{\theta}}{r}d\phi dt+r^2(d\theta^2+\sin^2{\theta} d\phi^2)
\end{equation}
and may be interpreted as the Schwarzschild metric with a nonstatic term.

In Refs. ~\cite{Brill:1966vsm,Lindblom:1974bq,PhysRev.173.1258} the problem of rotating shells was treated as a perturbation of the Schwarzschild geometry, instead of a Minkowski one. In this approximation \cite{Lindblom:1974bq,Brill:1966vsm}, an exact form was obtained for the ratio $\Omega/\omega$ (the dragging coefficient), where $\Omega=\Omega(r)$ is the angular velocity of the background  as observed by an observer at infinity and $\omega$ is the angular velocity of the spherical shell of radius $r_0$ relative to the flat infinity. Thus, Cohen~\cite{PhysRev.173.1258} investigating the gravitational collapse of rotating bodies, found that if the shell radius $r_0$ exceeds $(9/8)\times R_s$, where $R_s$ is the Schwarzschild radius, rotation can stabilize the shell and stop it from collapsing. On the other hand for a $R_s<r_0<\frac{9}{8} R_s$, the rotation can not stop the collapse of the shell.

An alternative approach to the stabilization of collapsing/expanding shells is associated with gravastars \cite{Lobo:2015lbc,Carter:2005pi}. These are structures conceived as non-singular  alternatives to black holes \cite{Mazur:2001fv, Antoniou:2020syc,Sakai:2014pga,Rahaman:2011hd}. A gravastar is a static spherically symmetric  Riemannian manifold, divided by a 3-dimensional surface in two pieces, namely in an exterior and an interior region separated by a ``gluing" surface at a radius close to that of Schwarzschild\cite{Cardoso:2007az}. The interior region, is a  de-Sitter space time with a fluid equation of state $p=-\rho$. This equation of state  violates the singularity theorem assumptions, thus producing a non-singular configuration. The outer region has a vacuum Schwarzschild geometry\cite{Cattoen:2006tk}, matched to the interior region through a (infinitesimally) thin shell with equation of state $p = \rho$ which helps to confine the interior region and ensures continuity of the metric. There have been several studies of gravastars that attempt to answer two basic questions:
\begin{itemize}
    \item Is a gravastar stable to generic perturbations?
    \item If stable can an observer distinguish a gravastar from a black hole of the same mass?
\end{itemize}

The Mazur-Mottola gravastar model is a spherically symmetric and static five-layer solution~\cite{Mazur:2001fv} to the Einstein equations including two infinitesimally thin shells needed by the junction conditions of the metric. The solution is thermodynamically stable but other stability analyses are not easy to perform because of the model's structure. Visser and Wiltshire (hereafter VW) have analyzed~\cite{Visser:2003ge} the radial stability of a gravastar with three layers. They found stable configurations and in addition, a number of bounded motion configurations obtained using the "bounded excursion conditions". In the case for proper initial conditions, the shell oscillates between two end points of an interval of the shell radius $R$ e.g.  $R\in(a_1, a_2)$, while for other initial conditions, the collapse (instability) occurs until the formation of a black hole.



Rotating gravastars have been analyzed in Ref.~\cite{Berti:2008xu} using the formalism of Refs. ~\cite{Hartle:1967he,Hartle:1968si,Friedman:1978hf}.  The properties of perturbed and rotating gravastars were considered in the limit of slow rotation approximation and their stability  against the ergoregion perturbations was investigated using the WKB approximation.  It was shown that most rotating gravastars are unstable even though, it was later shown that stable rotating gravastar configurations may also be found ~\cite{Berti:2008xu}.

The idea of shell stabilization using a deSitter interior solution has lead to the stable non-rotating gravastar solutions. In the gravastar case, the shell attractive gravity and tension are balanced by the repulsive gravity of the vacuum of the interior deSitter spacetime. The gravastar paradigm raises the following question: Can the repulsive properties of the vacuum gravastar interior be replaced by other more realistic mechanism with similar repulsive properties? In particular, can angular momentum of the interior spacetime play a stabilization role similar to the deSitter repulsive gravity role?  These questions are addressed in the present analysis.

In particular, we investigate in a qualitative and physical way the effects of rotation in the relativistic shells described in Refs. ~\cite{Israel:1966rt,Visser:2003ge}. We focus on slowly rotating configurations of mass $M$ and angular momentum $J$ such that $J/M^2<<1$ and address the following questions:
\begin{enumerate}
    \item Can angular momentum stabilize shell systems in the absence of a de Sitter interior?
    \item What are the conditions for this stabilization?
\end{enumerate}
In the context of addressing these questions we modify the analyses of Refs. ~\cite{Israel:1966rt,Visser:2003ge} by introducing both radial and angular perturbations. Using the original Israel conditions\cite{Israel:1966rt, Deruelle:2007pt}, we investigate the stability of rotating shells with a rotating interior spacetime.

The structure of this paper is the following: In  Section 2, we present the basic equations needed to discuss the stability of a rotating spherical shell which constitutes a boundary of  a Schwarzschild and/or Sitter space time. In Section 3, using the obtained shell dynamical equations we derive the evolution equation of the radius of a rotating spherical shell of radius $R$ and apply it in characteristic expanding models, assuming that the shell total mass remains constant and hence its surface density decreases like $\sim 1/R^2$ .  We thus derive new stability criteria in the context of a perturbative analysis. In section 4, we construct non-perturbative equations describing the energy conservation and the dynamics of a rotating shell. Stability criteria and bounded excursion stationary rotating shell configurations are also identified along the lines of corresponding gravastar solutions ~\cite{Visser:2003ge}. Finally in Section 5, we summarize our results and discuss possible extensions of the present work.

\section{Rotating Shell Dynamics and Israel Junction Conditions-The Equations}
We consider a 3-dimensional thin shell $\Sigma$, which separates a Riemannian space time into two 4-dimensional manifolds $V^{-}$, $V^{+}$. In the formalism of Refs. ~\cite{Israel:1966rt,Nakao:2018knn}, a space like unit vector $\bf{n}$, normal to the singular surface $\Sigma$, is directed from $V^{-}$ to $V^{+}$. Both regions are described by a single set of coordinates $x^{\mu}=(t,r,\theta,\phi)$ in which the line element reads~\cite{Israel:1966rt}
\begin{equation}\label{e1}
ds^2=-\big(1-\frac{2M(r)}{r}\big)dt^2+\frac{dr^2}{\big(1-\frac{2M(r)}{r}\big)}+r^2(d\theta^2+\sin^2{\theta}d\phi^2)
\end{equation}
For proper choices of $M(r)$, the metric (\ref{e1}) may describe either a Schwarzschild or a deSitter space time.

A perturbed rotating form of the metric (\ref{e1}) can describe the interior and the exterior metric  of a thin slowly rotating shell as suggested by Thirring~\cite{Thirring:1918uj, Brill:1966vsm} and adopted in Refs ~\cite{Lindblom:1974bq, DeLaCruz:1968zz}. This metric has the form 
\begin{equation}\label{e1a}
ds_{\pm}^2=-\Delta_{\pm} dt_{\pm}^2+\frac{dr_{\pm}^2}{\Delta_{\pm}}+r_{\pm}^2d\theta_{\pm}^2+r_{\pm}^2 \sin^2{\theta_{\pm}}[d\phi_{\pm}-\Omega_{\pm}(r) dt_{\pm}]^2,~~r_{+}> R,~~r_{-}<R
\end{equation}
On the shell, we have $R=r_{+}=r_{-}$, $\Delta_{\pm}=\big(1-\frac{2M(r_{\pm})}{r_{\pm}}\big)$ and  $\Omega_{\pm}=\Omega(r_{\pm})$ describes a local rotation of the inertial frames with respect to infinity.
If we set $\Omega_{\pm}=0$, we recover the nonrotating metric (\ref{e1}).


The evolution of a circular segment of the rotating shell is described by $x^{\mu}=(t(\tau), R(\tau),\theta=\theta_0, \phi(\tau))$, where the proper time $\tau$ of an observer lying on the surface $\Sigma$ of the shell is defined by
\begin{equation}\label{e3}
-d\tau^2=ds^2=-(\Delta-\Omega^2 R^2\sin^2{\theta}) dt^2+\frac{dR^2}{\Delta}-2\Omega R^2\sin^2{\theta} dt d\phi+R^2 d\theta^2+R^2 \sin^2{\theta}d\phi^2
\end{equation}
where $\Delta=(1-\frac{2 M(R)}{R})$.

From Eq.(\ref{e3}) we obtain $(\frac{d\tau}{dt})^2=\frac{W}{\chi}$, with $\chi\equiv \Delta+\dot{R}^2$ and $W\equiv \Delta[\Delta-R^2\sin^2{\theta}(\Omega-\omega)^2]$. Here, dot denotes the derivative with respect to $\tau$ and $\dot{\phi}\equiv\frac{d\phi}{d \tau}=\frac{d\phi}{dt}\frac{dt}{d\tau}=\omega\dot{t} $, with $\omega$ being the rotation rate of the shell. \footnote{Here,  $\omega$ is the angular velocity of the shell as seen by an observer at rest and in the slow rotation approximation~\cite{Chirenti:2008pf}, namely in  first order in the small parameter $\omega/\omega_k <1$, where $\omega_k$ is the Keplerian frequency (see Refs. \cite{Brill:1966vsm, Chirenti:2008pf, Cardoso:2007az}).}

For $\theta=\theta_0=const.$, the 4-velocity tangent to the two sides of the surface $\Sigma$ is given by the expression 
\begin{equation}\label{e11}
V_{\pm}^{\mu}=[\dot{t},\dot{R},\dot{\theta_0},\dot{\phi}]|_{\pm}\equiv[V^0, V^1,0, \omega V^0 ] |_{\pm}
\end{equation}
which satisfies the relation $V_{\mu} V^{\mu}|_{\pm}=-1$.

We now focus on the geometry~\cite{Lobo:2020kxn} on the singular surface $\Sigma$. This surface possesses 3-tangent vectors $e_{(i)}=\frac{\partial}{\partial \xi^i}|_{\pm}$ with components $e_{(i)}^{\mu}|_{\pm}=\frac{\partial x_{\pm}^{\mu}}{\partial\xi^{i}}$, where $\xi^{i}=(\tau,\theta,\phi)$. The explicit expressions of the $e_{(i)}^{\mu}|_{\pm}$ components are
\begin{eqnarray}\label{g1}
&&e_{(1)}^{\mu}|_{\pm}=[\dot{t},\dot{R},0,\omega \dot{t}]|_{\pm}\nonumber\\
&&e_{(2)}^{\mu}|_{\pm}=[0,0,1,0]|_{\pm}\nonumber\\
&&e_{(3)}^{\mu}|_{\pm}=[0,0,0,1]|_{\pm}
\end{eqnarray}
On the surface $\Sigma$, the basis (\ref{g1}) specifies the induced metric via the scalar  product
\begin{equation}\label{e2}
h_{ij}^{\pm}=e_{i}\cdot e_{j}|^{\pm}=g_{\mu\nu}e_{(i)}^{\mu}e_{(j)}^{\nu}|^{\pm}
\end{equation}
which explicitly reads 
\begin{equation}\label{g2}
ds_{\Sigma}^2=-d\tau^2+R^2 d\theta^2-2(\Omega-\omega) \dot{t} R^2 \sin^2{\theta}d \tau d\phi+R^2\sin^2{\theta} d\phi^2
\end{equation}

We assume that the metric (\ref{g2}) satisfies the first junction condition e.g, 
\begin{equation}\label{u1}
[[h_{\mu\nu}]]=0 
\end{equation}
where  $[[X]]\equiv X_{+}-X_{-}$ denotes the discontinuity in $X$ across the shell~\cite{Visser:2003ge, Lobo:2020kxn}. 
Eq.(\ref{u1}) is written as 
\begin{equation}\label{jcw1}
h_{\tau\phi}^{+}=h_{\tau\phi}^{-}\Rightarrow(\Omega_{+}-\omega)\dot{t}_{+}=(\Omega_{-}-\omega)\dot{t}_{-}
\end{equation}
Equation (\ref{jcw1}) implies that
\begin{eqnarray}\label{jw2}
&&(\Omega_{+}-\omega)\sqrt{\frac{\chi_{+}}{W_{+}}}=(\Omega_{-}-\omega)\sqrt{\frac{\chi_{-}}{W_{-}}}\Rightarrow\nonumber\\
&&(\Omega_{+}-\omega)\sqrt{\frac{\Delta_{+}+\dot{R}^2}{\Delta_{+}[\Delta_{+}-R^2\sin^2{\theta}(\Omega_{+}-\omega)^2]}}=(\Omega_{-}-\omega)\sqrt{\frac{\Delta_{-}+\dot{R}^2}{\Delta_{-}[\Delta_{-}-R^2\sin^2{\theta}(\Omega_{-}-\omega)^2]}}\nonumber\\
\end{eqnarray}
where $\dot{R}^2$ will be computed below from the second Israel junction condition (Lanczos equation). In what follows we assume slow rotation and ignore terms of powers higher than two of the angular velocities and thus of $(\Omega_{\pm}-\omega)$.

Thus, a further analysis of Eq. (\ref{jw2}) gives the approximate equation
\begin{eqnarray}\label{jw3}
&&\Delta_{-}(\Omega_{+}-\omega)\sqrt{\Delta_{+}+\dot{R}^2}[1-\frac{R^2\sin^2{\theta}}{2\Delta_{-}}(\Omega_{-}-\omega)^2]\nonumber\\
&&=\Delta_{+}(\Omega_{-}-\omega)\sqrt{\Delta_{-}+\dot{R}^2}[1-\frac{R^2\sin^2{\theta}}{2\Delta_{+}}(\Omega_{+}-\omega)^2]
\end{eqnarray}

Moreover, Eq. (\ref{jw3}) becomes
\begin{equation}\label{jc4}
\Delta_{-}(\Omega_{+}-\omega)\sqrt{\Delta_{+}+\dot{R}^2}
=\Delta_{+}(\Omega_{-}-\omega)\sqrt{\Delta_{-}+\dot{R}^2}
\end{equation}
or
\begin{equation}\label{jpp1}
(\Omega_{-}-\omega)=(\frac{\Delta_{-}}{\Delta_{+}})\sqrt{\frac{\Delta_{+}+\dot{R}^2}{\Delta_{-}+\dot{R}^2}}(\Omega_{+}-\omega)
\end{equation}
Thus, the requirement that the first junction condition $[[h_{\mu\nu}]]=0$ is valid, leads to the approximate Eq. (\ref{jpp1}) in which the $\dot{R}^2$ will be obtained with the aid of the second junction condition in subsection 3.1.  

In this context, we compute the extrinsic curvature given by the equation
\begin{equation}\label{k1}
K_{ij}^{\pm}=n_{(\mu;\nu)}e_{(i)}^{\mu} e_{(j)}^{\nu}|^{\pm}
\end{equation}
The normal vector to the 3-surface $\Sigma$ with normal condition $n_{\mu} n^{\mu}=1$, must satisfy the conditions $V^{\mu}n_{\mu}=0$,  $n_{\mu} e^{\mu}_{(i)}=0$, where $i=\tau,\theta,\phi$ and $\mu=t,r,\theta,\phi$. From these equations we find
\begin{eqnarray}\label{g6}
&&n_{0}=\pm\frac{\dot{R}}{\sqrt{g^{00}\dot{R}^2+g^{11}(\dot{t})^2}},~~n_1=\mp\frac{\dot{t}}{\sqrt{g^{00}\dot{R}^2+g^{11}(\dot{t})^2}}\nonumber\\
&&n_2=n_3=0
\end{eqnarray}
Using Eqs. (\ref{k1}) and (\ref{g6}) we compute the components of the extrinsic curvature $K_{ij}$, which are
\begin{equation}\label{g8ix}
K_{\tau\tau}=-n_0[ \ddot{t}+2\Gamma_{tr}^t\dot{t}\dot{R}+2\Gamma_{r\phi}^t\omega\dot{t}^2]
-n_1[\ddot{R}+(\Gamma_{tt}^r+2\omega\Gamma_{t\phi}^r+\omega^2\Gamma_{\phi\phi}^r)\dot{t}^2+\Gamma_{rr}^r\dot{R}^2]\nonumber\\
\end{equation}
\begin{eqnarray}\label{g8y}
K_{\theta\theta}&=&-n_1 \Gamma_{\theta\theta}^r,~~K_{\phi\phi}=\sin^2{\theta} K_{\theta\theta}\nonumber\\
K_{\tau\phi}&=&-n_0 \dot{R} \Gamma_{r\phi}^t -n_1 \dot{t} [\Gamma_{t\phi}^r+\omega \Gamma_{\phi\phi}^r]\nonumber\\
K_{\tau\theta}&=&0,~~\mbox{~and~}~~K_{\theta\phi}=0
\end{eqnarray}
Using these components of the extrinsic curvature $K_{ij}$, we may apply Israel's junction conditions~\cite{Israel:1966rt,Senovilla:2013vra} to relate the discontinuity in the extrinsic curvature to the surface stress-energy $S_{ij}$ located on the shell as follows~\cite{Israel:1966rt, Lanczos:1924zz} 
\begin{equation}\label{g11s}
S_{ij}=-\frac{1}{8\pi}\{[[K_{ij}-K h_{ij}]]\}
\end{equation}

Setting $Q_{ij}=[K_{ij}-K h_{ij}]$, Eq.(\ref{g11s}) is written as
\begin{equation}\label{g11a}
S_{j}^{i}=-\frac{1}{8\pi}[[Q^{i}_{j}]]
\end{equation}
Because of the symmetry of the metric (\ref{e1a}), the surface stress-energy tensor is non-diagonal and has the form
\begin{equation}\label{g11b}
 S_{j}^{i}=\left(
  \begin{array}{ccc}
    -\sigma & 0 & S_{\tau}^{\phi} \\
    0 & \theta_1 & 0 \\
    S_{\phi}^{\tau} & 0 & \theta_2 \\
  \end{array}
\right)=-\frac{1}{8\pi}\left(
          \begin{array}{ccc}
            [[Q_{\tau}^{\tau}]] & 0 & [[Q_{\tau}^{\phi}]] \\
            0 & [[Q_{\theta}^{\theta}]] & 0  \\
            ([Q_{\phi}^{\tau}]] & 0 & [[Q_{\phi}^{\phi}]] \\
          \end{array}
        \right)
\end{equation}
where $\theta_1, \theta_2$ are the surface tensions in the angular directions. 
The $Q_{j}^{i}$ components may be evaluated above ($(+)$) and below ($(-)$) at the shell radius. From Eq. (\ref{g11a}) we derive the shell dynamical equations
\begin{equation}\label{f1b}
4\pi\sigma R^2=-[[\{K_{\theta\theta}+\frac{\Omega^2}{2}(\frac{\chi}{W})K_{\theta\theta}-\frac{\Omega}{2}R^2(\sqrt{\frac{\chi}{W}})K_{\tau\phi}\}[1+\Omega^2R^2\sin^2{\theta}(\frac{\chi}{W})]^{-1}]]
\end{equation}
\begin{equation}\label{f2bx}
8\pi \theta_1 R^2=-[[\{R^2K_{\tau\tau}-K_{\theta\theta}+2\Omega R^2(\sqrt{\frac{\chi}{W}})K_{\tau\phi}\}[1+\Omega^2R^2\sin^2{\theta}(\frac{\chi}{W})]^{-1}]]
\end{equation}
and
\begin{eqnarray}\label{f3b}
&&8\pi \theta_2 R^2=\nonumber\\&&-[[\{R^2K_{\tau\tau}-K_{\theta\theta}+2\Omega R^2(\sqrt{\frac{\chi}{W}})K_{\tau\phi}-\Omega^2(\frac{\chi}{W})R^2\sin^2{\theta}K_{\theta\theta}\}[1+\Omega^2R^2\sin^2{\theta}(\frac{\chi}{W})]^{-1}]]\nonumber\\
\end{eqnarray}
Inserting the non-zero components of $K_{ij}$ into Eqs. (\ref{f1b})-(\ref{f3b}), we consider two characteristic shell models which are allowed to expand and to rotate slowly. We use the resulting equations to examine the shell stability in terms of the model parameters.
\section {Rotating Shell with a Schwarzchild Exterior Spacetime}
In this section, we consider the special, but physically interesting case, where there is a Schwarzchild exterior spacetime while the shell and the interior are allowed to rotate.

In this model we set $\Omega_{+}=0$, $M_{+}=M=const.$, $M_{-}=M(R)=kR^3$, with $k\in[0,1)$ and $\omega$ is proportional to Keplerian frequency $\omega_k<1$.
The Eqs. (\ref{f1b}), (\ref{f2bx}) and (\ref{f3b}) become
\begin{eqnarray}\label{x1}
&&4\pi\sigma R^2[1+R^2\sin^2{\theta}\Omega_{-}^2\frac{\chi_{-}}{W_{-}}]\nonumber\\
&&=K_{\theta\theta}^{+}-K_{\theta\theta}^{-}\nonumber\\
&&+\frac{1}{2}R^2\Omega_{-}\sqrt{\frac{\chi_{-}}{W_{-}}} K_{\tau\phi}^{-}-\Omega_{-}^2 R^2\sin^2{\theta}(\frac{\chi_{-}}{W_{-}})[\frac{1}{2}K_{\theta\theta}^{-}- K_{\theta\theta}^{+}]
\end{eqnarray}
\begin{eqnarray}\label{x2}
&&-8\pi \theta_1 R^2\{1+R^2\sin^2{\theta}\frac{\Omega_{-}^2\chi_{-}}{W_{-}}\}\nonumber\\
&&=[R^2K_{\tau\tau}^{+}-K_{\theta\theta}^{+}]-[R^2 K_{\tau\tau}^{-}-K_{\theta\theta}^{-}]-2R^2[\Omega_{-}\sqrt{\frac{\chi_{-}}{W_{-}}} K_{\tau\phi}^{-}]\nonumber\\
&&+R^2\sin^2{\theta}\{\Omega_{-}^2\frac{\chi_{-}}{W_{-}}[R^2 K_{\tau\tau}^{+}-K_{\theta\theta}^{-}]\}
\end{eqnarray}
and
\begin{eqnarray}\label{x3}
&&-8\pi \theta_2 R^2\{1+R^2\sin^2{\theta}[\frac{\Omega_{-}^2 \chi_{-}}{W_{-}}]\}\nonumber\\
&&=[R^2K_{\tau\tau}^{+}-K_{\theta\theta}^{+}]-[R^2 K_{\tau\tau}^{-}-K_{\theta\theta}^{-}]+R^2\sin^2{\theta}[\Omega_{-}^2\frac{\chi_{-}}{W_{-}} K_{\theta\theta}^{-}]\nonumber\\
&&-2R^2[\Omega_{-}\sqrt{\frac{\chi_{-}}{W_{-}}} K_{\tau\phi}^{-}]\nonumber\\
&&+R^2\sin^2{\theta}\{\Omega_{-}^2\frac{\chi_{-}}{W_{-}}[R^2 K_{\tau\tau}^{+}-K_{\theta\theta}^{+}]]\}
\end{eqnarray}
We now focus on Eq. (\ref{x1}), which is the main equation. In the context of the slow rotation assumption, we expand Eq. (\ref{x1}) in powers of $\Omega_{-}$, $(\Omega_{-}-\omega)$,  we ignore terms higher than two and  keep the signs $\pm$ where needed.
Inserting the non-zero $K_{ij}$ of Eqs. (\ref{g8ix}) into Eq. (\ref{x1}), we derive the following approximate equation
\begin{eqnarray}\label{y3}
&&8\pi\sigma R^2[1-\dot{R}^2\frac{R^2\sin^2{\theta}}{2\Delta_{+}^2}\omega^2]\sqrt{\chi_{+}}\nonumber\\
&&=R^3\sin^2{\theta}[p_4\dot{R}^4+p_2\dot{R}^2+p_0]+s_0
\end{eqnarray}
where
\begin{eqnarray}\label{y2}
&&s_0=-16\pi^2\sigma^2 R^3+2[M_{+}-M_{-}]=-\frac{M_s^2}{R}+2[M_{+}-M_{-}],~~\mbox{where~}~~M_s=4\pi\sigma R^2\nonumber\\
&&p_4=[\frac{\omega^2}{\Delta_{+}^2}-\frac{(\Omega_{-}-\omega)^2}{\Delta_{-}^2}-\frac{\omega\Omega_{-}}{\Delta_{-}^2}]\nonumber\\
&&p_2=[\frac{\omega^2}{\Delta_{+}}-\frac{(\Omega_{-}-\omega)^2}{\Delta_{-}}-2\frac{\omega\Omega_{-}}{\Delta_{-}}+\frac{R\Omega_{-}(\Omega_{-})_{,R}}{2\Delta_{-}}]\nonumber\\
&&p_0=[\frac{1}{2}R\Omega_{-}(\Omega_{-})_{,R}-\omega\Omega_{-}]
\end{eqnarray}
We square both sides of Eq. (\ref{y3}) and after some elementary calculations we end up with the equation
\begin{equation}\label{w1}
Q_4\dot{R}^4+Q_2\dot{R}^2+Q_0=0,~~\mbox{~where~}~~Q_4\neq 0
\end{equation}
where
\begin{eqnarray}\label{w2}
Q_4&=&R^3\sin^2{\theta}\{[M_{+}-M_{-}-\frac{M_s^2}{2 R}]p_4+\frac{M_s^2}{R\Delta_{+}^2}\omega^2\}\nonumber\\
Q_2&=&-M_s^2+R^3\sin^2{\theta}\{\frac{M_s^2}{\Delta_{+}}\omega^2+[M_{+}-M_{-}-\frac{M_s^2}{2R}]p_2\}\nonumber\\
Q_0&=&R^3\sin^2{\theta}[M_{+}-M_{-}-\frac{M_s^2}{2R}]p_0-\{M_s^2+4M_{+}M_{-}-[\frac{M_s^2}{2R}+(M_{+}+M_{-})]^2\}\nonumber\\
\end{eqnarray}
It easy to verify that for $\Omega_{\pm}=0,\omega=0$, Eq. (\ref{w1}) reduces to Eq. (40) of Ref.~\cite{Visser:2003ge} as expected. Also, for $\Omega_{\pm}=\omega=0$, $M_s=const.$ and $M_{-}=0$, we recover Israel's equation (43) of Ref.~\cite{Israel:1966rt} provided that the parameter "$a$", appearing in this equation, is $ a =M/M_s$ and the surface density $\sigma$ is proportional to $1/R^2$.
Eq. (\ref{w1}) for $\Omega_{\pm}\neq 0$ and $ \omega\neq 0$ leads to the shell radius velocity
\begin{equation}\label{ww2}
\dot{R}_{1,2}^2=\frac{1}{2Q_4}\bigg(-Q_2\pm\sqrt{Q_2^2-4 Q_4 Q_0}\bigg)
\end{equation}
which is a zero energy like equation of the form $\frac{1}{2}\dot{R}^2 +V_{1,2}(R)=0$ with
\begin{equation}\label{ww3}
V_{1,2}(R)=-\frac{1}{4Q_4}\bigg(Q_2\mp\sqrt{Q_2^2-4 Q_4 Q_0}\bigg)
\end{equation}
So far we have:(i) From the first junction condition we obtained the approximate Eq.(\ref{jpp1}).
(ii) From the second junction condition we derived the Eqs.(\ref{x1})-(\ref{x3}) and from the Eq.(\ref{x1}) we had two expressions for the shell radius velocity $\dot{R}^2$. Setting one of those expressions of $\dot{R}^2$, for example the $\dot{R_{1}}^2$, into Eq.(\ref{jpp1}) we obtain a rather complicated equation which is simplified in what follows in the context of a slow rotation perturbative expansion.

\subsection{Stability of a shell configuration}
We now assume a deSitter interior (gravastar) and introduce a perturbative (slow rotation) expansion in Eq. (\ref{w1}). We thus decompose $\dot{R}^2$ and $Q_i$ as follows
\begin{equation}\label{d1px}
\dot{R}^2=\dot{R}_{(0)}^2+\dot{R}_{(2)}^2
\end{equation}
and
\begin{equation}\label{w5px}
Q_i=Q_i^{(0)}+Q_i^{(2)},~~i=0,2,4
\end{equation}
where the quantities $\dot{R}_{(0)}^2$, $Q_i^{(0)}$ are independent of $\Omega_{\pm}$ and $\omega$, while the quantities $\dot{R}_{(2)}^2$, $Q_i^{(2)}$ (here $i=0,2,4$) include terms proportional to $(\Omega_{\pm}-\omega)^2$ and $\omega^2$. Inserting Eqs. (\ref{d1px}) and (\ref{w5px}) into equation Eq. (\ref{w1}) we obtain
\begin{equation}\label{w6px}
Q_4^{(2)}[\dot{R}_{(0)}^4+\dot{R}_{(2)}^4+2\dot{R}_{(0)}^2\dot{R}_{(2)}^2]+[Q_2^{(0)}+Q_2^{(2)}][\dot{R}_{(0)}^2+\dot{R}_{(2)}^2]+[Q_0^{(0)}+Q_0^{(2)}]=0
\end{equation}
The function $Q_4$ is of second order i.e.  $Q_4=Q_4^{(2)}$ because it includes terms of second order in $\Omega_{\pm}$ and $\omega$.
The quantities $Q_2$ and $Q_0$ are of the form
\begin{eqnarray}\label{w3p}
&&Q_2\equiv Q_2^{(0)}+Q_2^{(2)}= -M_s^2+Q_2^{(2)}\nonumber\\
&&Q_0\equiv Q_0^{(0)}+Q_0^{(2)}=-\{M_s^2+4M_{+}M_{-}-[\frac{M_s^2}{2R}+(M_{+}+M_{-})]^2\}+Q_0^{(2)}
\end{eqnarray}
where
\begin{eqnarray}\label{w4p}
&&Q_4^{(2)}=R^3\sin^2{\theta}\{[M_{+}-M_{-}-\frac{M_s^2}{2 R}]p_4+\frac{M_s^2}{R\Delta_{+}^2}\omega^2\}\nonumber\\
&&Q_2^{(2)}=+R^3\sin^2{\theta}\{\frac{M_s^2}{\Delta_{+}}\omega^2+[M_{+}-M_{-}-\frac{M_s^2}{2R}]p_2\}\nonumber\\
&&Q_0^{(2)}=R^3\sin^2{\theta}[M_{+}-M_{-}-\frac{M_s^2}{2R}]p_0
\end{eqnarray}


In Eq. (\ref{w6px}), the terms $Q_4^{(2)} \dot{R}_{(2)}^4$, $2Q_4^{(2)}\dot{R}_{(0)}^2\dot{R}_{(2)}^2$ and $Q_2^{(2)}\dot{R}_{(2)}^2$ are of higher order than two (we ignore these terms). The obtained approximate equation now reads
\begin{equation}\label{w7px}
Q_4^{(2)}\dot{R}_{(0)}^4+Q_2^{(0)} \dot{R}_{(0)}^2+Q_2^{(2)}\dot{R}_{(0)}^2+Q_2^{(0)}\dot{R}_{(2)}^2+Q_0^{(0)}+Q_0^{(2)}=0
\end{equation}
Further, from Eq. (\ref{w7px}) we have:

{\bf Zero order equation:}
\begin{eqnarray}\label{w7ap}
\dot{R}_{(0)}^2&=&-\frac{Q_0^{(0)}}{Q_2^{(0)}}\nonumber\\
&=&-\frac{1}{M_s^2}\{M_s^2+4M_{+}M_{-}-[\frac{M_s^2}{2R}+(M_{+}+M_{-})]^2\}
\end{eqnarray}

{\bf Second order:}
\begin{equation}\label{w8p}
Q_4^{(2)} \dot{R}_{(0)}^4+Q_2^{(0)}\dot{R}_{(2)}^2+Q_2^{(2)}\dot{R}_{(0)}^2+Q_0^{(2)}=0
\end{equation}
which becomes
\begin{equation}\label{w9p}
\dot{R}_{(2)}^2=-\frac{1}{Q_2^{(0)}}[Q_4^{(2)} \dot{R}_{(0)}^4+Q_2^{(2)}\dot{R}_{(0)}^2+Q_0^{(2)}]
\end{equation}

The equation (\ref{jpp1})  emerges by imposing continuity of the  induced metric  i.e. $[[h_{\mu\nu}]]=0$. Decomposing Eq. (\ref{jpp1}) in the context of the slow rotation expansion, we express $\dot{R}^2$ as in Eq.(\ref{d1px}) to obtain to first order in $(\Omega_{-}-\omega)$
\begin{equation}\label{ppf}
(\Omega_{-}-\omega)=(\frac{\Delta_{-}}{\Delta_{+}})\sqrt{\frac{\Delta_{+}+\dot{R}_{(0)}^2}{\Delta_{-}+\dot{R}_{(0)}^2}}(\Omega_{+}-\omega)
\end{equation}
Since we require an exterior Schwarzschild space-time,  we set $\Omega_{+}=0$ and thus, Eq. (\ref{ppf}) becomes
\begin{equation}\label{pp}
\Omega_{-}=\omega [1-(\frac{\Delta_{-}}{\Delta_{+}})\sqrt{\frac{\Delta_{+}+\dot{R}_{(0)}^2}{\Delta_{-}+\dot{R}_{(0)}^2}}]
\end{equation}
In the context of a Schwarzschild exterior and a deSitter interior geometry, we have $\Delta_{+}=1-\frac{2M_{+}}{R}$, $M_{+}=M=const.$, $\Delta_{-}=1-\frac{2M_{-}}{R}=1-2k R^2$ and $M_s=4\pi\sigma R^2$. This formulation leads to  
\begin{equation}\label{pp3}
Z\equiv (\frac{\Delta_{-}}{\Delta_{+}})\sqrt{\frac{\Delta_{+}+\dot{R}_{(0)}^2}{\Delta_{-}+\dot{R}_{(0)}^2}}=\frac{R(1-2kR^2)}{R-2M}[R^3-\frac{M}{k+8\pi^2\sigma^2}][R^3+\frac{M}{k-8\pi^2\sigma^2}]^{-1}
\end{equation}
where in the last equality we have used Eq. (\ref{w7ap}).
\section{Rotating spherical shell: Energy conservation and dynamics}
The dynamics and energy conservation of a rotating spherical dust shell may be studied using the Gauss-Codazzi equations (see Refs. ~\cite{Lobo:2020kxn, Lanczos:1924zz, DeLaCruz:1970kk,Yamanaka:1992zd} and Refs. in ~\cite{Yamanaka:1992zd})
\begin{eqnarray}\label{new1x}
&&^{(3)}R+\tilde{K}_{\mu\nu} \tilde{K}^{\mu\nu}-\tilde{K}^2=-16\pi^2G^2(S_{\mu\nu} S^{\mu\nu}-\frac{1}{2} S^2)-8\pi G\{T_{\Sigma}^{\mu\nu} n_{\mu}n_{\nu}\}\\
&&D_{\nu}\tilde{K}_{\mu}^{\nu}-D_{\mu}\tilde{K}=4\pi G \{T_{\Sigma}^{\rho\sigma} n_{\rho}h_{\mu\sigma}\}
\end{eqnarray}
along with the Israel (Lanczos) Eq.(\ref{g11s}) and 
Einstein's field equations
with their projections 
\begin{eqnarray}\label{new2}
G_{\rho\sigma}n^{\rho}n^{\sigma}&=&8\pi G T_{\rho\sigma}n^{\rho} n^{\sigma}\nonumber\\
G_{\rho\sigma}n^{\rho}h_{\mu}^{\sigma}&=&8\pi G T_{\rho\sigma}n^{\rho} h_{\mu}^{\sigma}
\end{eqnarray}
From the above Eqs. (\ref{new2}) and  Eqs. (\ref{g6}), (\ref{g2}), the following relations may be obtained
\begin{eqnarray}\label{new3}
&&\tilde{K}_{\mu\nu} S^{\mu\nu}=[[T_{\Sigma}^{\mu\nu} n_{\mu}n_{\nu}]]\\
\label{new3a}
&&D_{\nu} S_{\mu}^{\nu}=-[[T_{\Sigma}^{\rho\sigma} n_{\rho}h_{\mu\sigma}]]
\end{eqnarray}
where $D_{\nu}$ is the covariant derivative in terms of the intrinsic metric (\ref{g2}) on $\Sigma$ and $\tilde{K}=K^{+}+K^{-}$. The Eqs. (\ref{new3}), (\ref{new3a}) are the Hamiltonian and momentum constrains, respectively.

For a dust shell, the energy momentum tensor is $T^{\mu\nu}=\sigma V^{\mu} V^{\nu}$, where the components of the 4-velocity are given by Eq. (\ref{e11}). In this case, the right hand sides of Eqs. (\ref{new3}), (\ref{new3a}) are identically zero. Equation (\ref{new3a})
describes the energy conservation of $S_{\mu}^{\nu}$ and leads to 
\begin{equation}\label{new5}
-\dot{\sigma}-\sigma \Gamma_{i\tau}^{i}+S_{\tau,\phi}^{\phi}+S_{\tau}^{\phi} \Gamma_{i\phi}^{i}=0
\end{equation}
where $\Gamma_{i\tau}^{i}=\frac{1}{2} h^{ij} h_{ij,\tau}$. Equation (\ref{new5}) becomes
\begin{equation}\label{new9a}
\frac{\dot{\sigma}}{\sigma}+2\frac{\dot{R}}{R}=\Pi\dot{R}\Rightarrow \sigma'=-\frac{2}{R}+\Pi
\end{equation}
where $\sigma'=\frac{d\sigma}{dR}$, 

\begin{equation}\label{new9ap}
\Pi=-(\Omega-\omega)R^2\sin^2{\theta}\frac{\sqrt{\chi_0}}{\Delta}\dot{R}_{(0)}[(\frac{\sqrt{\chi_0}}{\Delta})_{,R}(\Omega-\omega) -(\frac{\sqrt{\chi_0}}{R\Delta})(\Omega+\omega)]
\end{equation}
and $\chi_0=\Delta +\dot{R}_{(0)}^2$.

Equation (\ref{new9a}) may be written as
\begin{equation}\label{new9b}
\frac{d (A\sigma)}{A d\tau}=-(\Omega-\omega)R^2\sin^2{\theta}\frac{\sqrt{\chi_0}}{\Delta}\dot{R}_{(0)}[(\frac{\sqrt{\chi_0}}{\Delta})_{,R}(\Omega-\omega) -(\frac{\sqrt{\chi_0}}{R\Delta})(\Omega+\omega)]
\end{equation}
where $A=4\pi R^2$ . In the case where $\omega=\Omega=0$, Eq. (\ref{new9b}) reduces to 
\begin{equation}\label{new7}
M_s=4\pi\sigma R^2=const.
\end{equation}
as expected for a dust shell. Equation (\ref{new9b}) implies that the rotation of the thin dust sell induces a nonzero dragging effect. In addition,  the internal energy of the rotating shell is not conserved, in contrast to the case of non-rotating shell ($\omega=0$ and  $\Omega=0$).

The dynamics of the shell may be obtained from Eq. (\ref{new3}), whose right hand side is identically zero, while its left hand side may be written as 
\begin{eqnarray}\label{new19x}
\tilde{K}_{\tau\tau}\sigma+\frac{\theta_1+\theta_2}{R^2}\tilde{K}_{\theta\theta}&=&-(\{\frac{1}{\sqrt{\chi_{+}}}[\ddot{R}-(\frac{M_{+}}{R})_{,R}]+\frac{1}{\sqrt{\chi_{-}}}[\ddot{R}-(\frac{M_{-}}{R})_{,R}]
\}\{-\Omega^2R^2\sin^2{\theta}\frac{\chi}{\Delta^2}\nonumber\\
&+&\Omega R^2\sin^2{\theta}\frac{\sqrt{\chi}}{\Delta^2}[\frac{(2\Omega-\omega)\chi}{R}+\frac{\Delta Q_{,R}}{2}]\}\nonumber\\
&+&R\sin^2{\theta}\sqrt{\chi_{+}}\{-\frac{\Omega\sqrt{\chi}}{\Delta}[(-\frac{\Omega}{\Delta})(\ddot{R}-(\frac{M}{R})_{,R})+\frac{(\Omega-\omega)\chi}{R}+\frac{\Delta\Omega_{,R}}{2}]\nonumber\\
&+&(\frac{\Omega^2}{8\pi}\frac{\chi}{\Delta^2})([[\frac{1}{\sqrt{\chi}}[\ddot{R}-(\frac{M}{R})_{,R}]+\frac{\sqrt{\chi}}{R}]]) \}\nonumber\\
&+&\frac{R^2\sin^2{\theta}}{\Delta_{-}}[\frac{\Delta_{-}(\Omega_{-})_{,R}}{2}-\frac{(\Omega_{-}-\omega)\chi_{-}}{R}]\{\frac{\Omega}{\Delta}[\ddot{R}-(\frac{M}{R})_{,R}\nonumber\\
&-&\frac{(\Omega-\omega)\chi}{R}-\frac{\Delta\Omega_{,R}}{2}+\Omega\sigma\frac{\sqrt{\chi}}{2}\nonumber\\
&+&(\frac{\Omega}{8\pi}\frac{\sqrt{\chi}}{4\Delta})([[\frac{1}{\sqrt{\chi}}[\ddot{R}-(\frac{M}{R})_{,R}]+\frac{\sqrt{\chi}}{R}]])\nonumber\\
&+&\frac{R^2\sin^2{\theta}}{\Delta}[\frac{(\Omega-\omega)\chi}{R}+\frac{\Delta\Omega_{,R}}{2}]\}
\end{eqnarray}
where
\begin{equation}\label{new17}
\theta_2\simeq \theta_1+\frac{\sin^2{\theta}}{8\pi}[[\Omega^2\frac{\chi}{\Delta^2}K_{\theta\theta}]]
\end{equation}
and $\theta_1$ (the shell's stress in the polar direction) is given by
\begin{eqnarray}\label{vp2}
-8\pi\theta_1 R^2&=&[[\frac{R^2}{\sqrt{x}}[\ddot{R}-(\frac{M}{R})_{,R}]+R\sqrt{\chi}]]-[[\Xi]]\nonumber\\
&-&2\Omega_{-}R^3\sin^2{\theta}(\frac{\sqrt{\chi_{-}}}{2\Delta_{-}^2})[R\Delta_{-}(\Omega_{-})_{,R}+2\chi_{-}(\Omega_{-}-\omega)]\nonumber\\
&-&(\Omega_{-}-\omega)^2 R^2\sin^2{\theta}(\frac{\chi_{-}}{\Delta_{-}^2})\{\frac{R^2}{\sqrt{\chi_{+}}}[\ddot{R}-(\frac{M}{R})_{,R}]+R\sqrt{\chi_{+}}\}
\end{eqnarray}
with
\begin{eqnarray}\label{vn1}
\Xi&\equiv &\frac{\sqrt{\chi} R^3 \sin^2{\theta}}{\Delta}[(\frac{M}{R})_{,R}R(\Omega-\omega)^2+\Delta(\Omega-\omega)^2+\Delta R\Omega_{,R}(\Omega-\omega)]\nonumber\\
&+&\dot{R}^2\frac{R^4\sin^2{\theta}}{2\sqrt{\chi}\Delta^2}[\ddot{R}-(\frac{M}{R})_{,R}](\Omega-\omega)^2+R\sqrt{\chi}[\dot{R}^2\frac{R^2\sin^2{\theta}}{2\Delta^2}(\Omega-\omega)^2]
\end{eqnarray}

As expected, from  Eq. (\ref{new19x}) with $\Omega=0$, $\omega=0$, we have $\theta_1=\theta_2=\theta_0$. For a dust shell Eq. (\ref{new19x}) reduces to
\begin{equation}\label{new18b}
\sigma\tilde{K}_{\tau\tau}+\frac{2\theta_0}{R^2}\tilde{K}_{\theta\theta}=0
\end{equation}
with $\theta_0=0$ (dust shell).
If the thin dust shell does not rotate, the solution of  Eq.  (\ref{new18b}) can in principle lead to the evolution of the shell radius $R(t)$. 

In the case where the shell rotates,  the equations (\ref{new9a}) and (\ref{new19x}) may lead to expressions for  $\sigma=\sigma(R,\Omega,\omega)$ and to the shell radius evolution $R(t)$ for given shell equations of state $\theta_1(\sigma)$, $\theta_2(\sigma)$.

\subsection{ Stability criteria}
The stability criteria of thin shells with no-rotation has been examined in Refs.~\cite{Chirenti:2008pf, Rosa:2020hex, LeMaitre:2019xez, Sharif:2020nys}. In this work, we generalize these equations to the rotating dust thin shell in backgrounds given by Eq. (\ref{e1a}). 

We re-write Eq. (\ref{w1}) with $\Omega_{\pm}\neq 0, \omega\neq 0$ as
\begin{equation}\label{w1a}
\dot{R}^4+P_2(R)\dot{R}^2+P_0(R)=0
\end{equation}
where $P_2(R)=\frac{Q_2}{Q_4}$ and $P_0(R)=\frac{Q_0}{Q_4}$.
Equation (\ref{w1a}) may be viewed as a generalized (zero) energy conservation and admits the solution $\dot{R_0}=0$ (static case) provided that $P_0(R_0)=0\Rightarrow Q_0(R_0)=0$.
By considering a perturbed configuration $R=R_0+\delta R(t)$, it is easy to show that the configuration $R=R_0$  is stable (the perturbation does not grow exponentially with time) if the 'energy' is bounded from below at zero, i.e if the following conditions are satisfied
\begin{equation}\label{w1d}
P_0(R)|_{R=R_0}=0,~~P_2(R)|_{R=R_0}>0,~~\frac{d P_0}{d R}|_{R=R_0}=0,~~\frac{d^2 P_0}{d R^2}|_{R=R_0}>0
\end{equation}
or in terms of $Q_0, Q_2, Q_4$
\begin{eqnarray}\label{w1e}
&&Q_0(R)|_{R=R_0}=0,~~\frac{d Q_0}{d R}|_{R=R_0}=0,~~\frac{d^2 Q_0}{d R^2}|_{R=R_0}>0 ~~\mbox{~and~} Q_4(R)|_{R=R_0}>0, Q_2(R)|_{R=R_0}>0~~\mbox{or~}\nonumber\\
&&Q_0(R)|_{R=R_0}=0,~~\frac{d Q_0}{d R}|_{R=R_0}=0,~~\frac{d^2 Q_0}{d R^2}|_{R=R_0}<0 ~~\mbox{~and~} Q_4(R)|_{R=R_0}<0, Q_2(R)|_{R=R_0}<0\nonumber\\
\end{eqnarray}
Equations (\ref{w1e}) constitutes a criterion for the identification of stability of rotating thin shells. 

In addition to static stable rotating shell configurations, oscillating and rotating 'bounded excursion' shell configurations may also be considered in analogy with the non-rotating corresponding gravastar configurations of Ref. \cite{Visser:2003ge}. In the 'bounded excursion' context, there exist two radii $R_1$, $R_2$ with restoring forces on the shell such that 
\begin{equation}\label{l1}
V(R)_{R=R_1}=0,~~\frac{dV}{dR}|_{R=R_1}\leq 0,~~V(R)_{R=R_2}=0,~~\frac{dV}{dR}|_{R=R_2}\geq 0
\end{equation}
with $V(R)<0$ for every $R\in(R_1,R_2)$.

\begin{figure}[h!]
\centering
\vspace{0cm}\rotatebox{0}{\vspace{0cm}\hspace{0cm}\resizebox{0.9\textwidth}{!}{\includegraphics{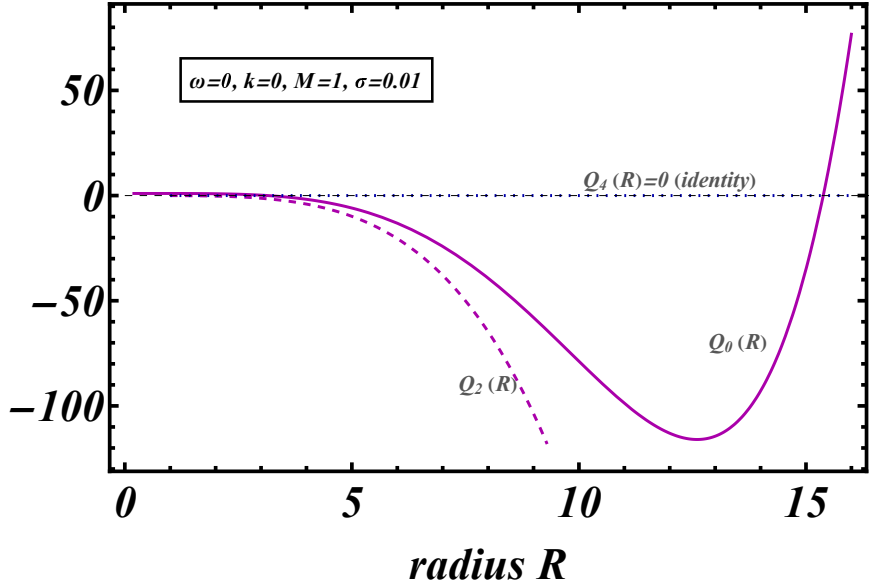}}}
\caption{The functions $Q_0(R)$, $Q_2(R)$ and $Q_4(R)$ for $k=0$, $\omega=0$ and $\sigma=const.$. In this case we have $Q_4(R)=0$ and $Q_2(R)<0$. In this configuration, the criteria (\ref{l2}) are not satisfied  and the model is unstable. }\label{fig1}
\end{figure}

\begin{figure}[h!]
\centering
\vspace{0cm}\rotatebox{0}{\vspace{0cm}\hspace{0cm}\resizebox{0.9\textwidth}{!}{\includegraphics{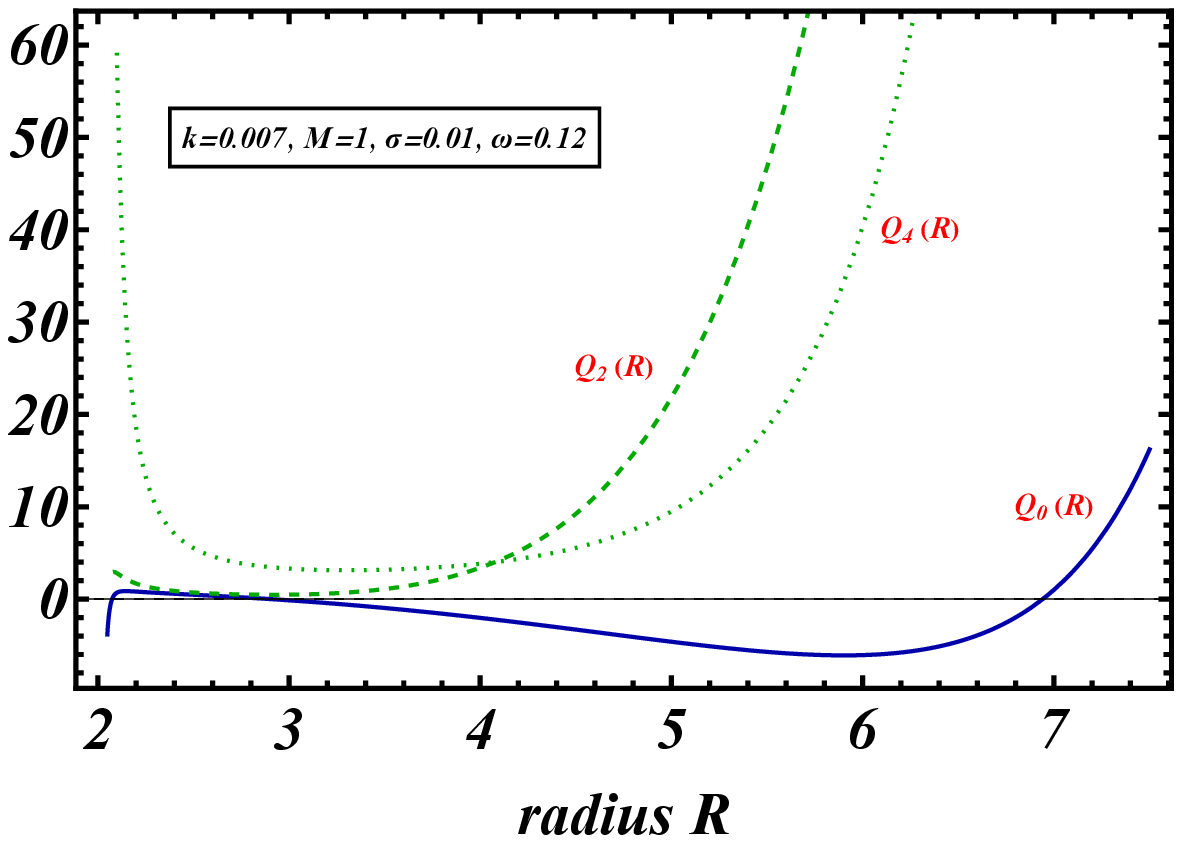}}}
\caption{The functions $Q_0$, $Q_2$ and $Q_4$ versus radius R (in Schwarzschild radii). The criterion (\ref{l2}) is satisfied and the configuration is stable in the interval $R\in (2.9 - 7.0)$. The minimum of $Q_0$ exists at $R_0\simeq 5.9 R_s$.}\label{fig2}
\end{figure}

Thus,  eqs. (\ref{w1e}) and (\ref{l1}) lead to a rather generalized 'bounded excursion' existence criterion of the form 
\begin{eqnarray}\label{l2}
&&Q_0(R)|_{R=R_1}=0,~~\frac{dQ_0(R)}{d R}|_{R=R_1}<0\nonumber\\
&&Q_0(R)|_{R=R_2}=0,~~\frac{dQ_0(R)}{d R}|_{R=R_2}>0,~~\mbox{~and~}\nonumber\\
&&Q_2(R)>0~~Q_4(R)>0, ~~\mbox{~for~every~}~ R\in(R_1,R_2)
\end{eqnarray}
In Fig. \ref{fig1} we show an example of a set of parameters that lead to to a non-rotating shell with no deSitter interior ($k=0$) and thus correspond to an unstable configuration ($P_2(R_0)<0$) as expected. In Fig. \ref{fig2} we show a set of parameters that lead to to a 'bounded excursion' rotating shell configuration with deSitter interior. In this case, the criterion (\ref{l2}) is satisfied while the full stability criterion (\ref{w1d}) is not satisfied.

There are three main factors that determine the qualitative behavior of the evolution of the shell.
\begin{itemize}
    \item {\bf The equation of state of the fluid inside the shell:} A fluid with equation of state parameter $w<-1/3$ induces repulsive gravitational effects which tend to counterbalance the tension and gravitational effects of the shell that favor collapse.  Thus, a fluid with $w<-1/3$ in the interior can lead to stabilization of the shell. This is the case of gravastars where the vacuum ($w=-1$), which dominates in the interior, stabilizes the shell.
    \item{\bf The angular momentum of the shell and/or the interior fluid:} The centrifugal effects induced angular momentum can also counterbalance the tension and gravitational effects of the shell that favor collapse thus leading to stabilization. This has been demonstrated in the present analysis.
    \item{\bf The equation of state of the shell fluid:} A shell fluid with positive equation of state parameters  has positive pressure that favors expansion and can balance the gravitational effects induced by a possible black hole in the center of the shell. In contrast, a vacuum $w=-1$ shell has negative pressure (tension) and therefore needs to be balanced by either an angular momentum (point 2 above) or by repulsive gravitational effects of the interior fluid (point 1 above).
    
    While for a rotating dust shell energy conservation is expressed by Eq. (\ref{new9b}), in the absence of rotation but for a general shell ($\theta_0\neq 0$), energy conservation on the shell implies that
    \begin{equation}\label{c1r}
    \frac{d}{d\tau}\big(\sigma R^2\big)-\theta_0 \frac{dR^2}{d\tau}=0
    \end{equation} 
    (see  Eq. (2.15) in Ref. ~\cite{Alestas:2020wwa}), which for a shell fluid equation of state $\theta_0=-w \rho$ leads to 
    \begin{equation}\label{c2r}
    \sigma=\sigma_0(\frac{R}{R_0})^{-2(w+1)}
    \end{equation}
    (see Eq.(2.20) in Ref. ~\cite{Alestas:2020wwa}).  Thus, we conclude that $\sigma \sim R^{-2w-2}$ and the total energy of the shell varies with  $R$ as $E\sim \sigma R^2 \sim R^{-2w}$. Ignoring the effects of gravity, for $w>0$ (positive pressure) the energy decreases with $R$ and the expansion of the shell is favored (the shell has pressure). In contrast, for $w<0$ (negative pressure) the energy increases with $R$ and thus contraction is favored (the shell has tension).
\end{itemize}

\section{Conclusions-Discussion}
We have investigated the stability of generalized relativistic slowly rotating shell configurations allowing for the following degrees of freedom (metric parameters): Central point mass, deSitter interior, angular momentum of the interior/exterior manifold and the shell equation of state. We have thus derived criteria for static-stable or oscillating bounded excursion rotating shell configurations demonstrating that such configurations exist even without the repulsive gravity effects of an interior deSitter spacetime. We have attributed the existence of such shell configurations to the repulsive effects of angular momentum that may replace the repulsive stabilizing effects of an interior deSitter space (appearing in the case of gravastars) and stabilize the shell against the attractive gravitational effects induced by the shell and central masses.

In particular we focused on relativistic rotating thin shells with angular velocity $\omega$ in a rotating background with $\Omega_{\pm}(R)$ with respect to spatial infinity. We presented the dynamical equations of the rotating shell configuration given by Eqs. (\ref{f1b})-(\ref{f3b}) allowing also for evolution of the radius of the shell. Even though previous works have considered separately rotation \cite{Uchikata:2016qku} or radial evolution of the shell \cite{Visser:2003ge}, our approach allows for both rotation and radial evolution/perturbation. 

The shell configurations considered  have three parts (layers): an exterior Schwarzschild spacetime, an interior Schwarzschild de Sitter spacetime and in the space between them, a thin rotating shell with surface density $\sigma$ and surface tensions $\theta_{1}$, $\theta_{2}$  and mass $M_s$.

We found a general dynamical Eq. (\ref{w1}), which allows an interior deSitter spacetime through the parameter $k$, expressing the curvature of the interior geometry due to  vacuum energy~\cite{Carter:2005pi}. It also includes the angular momentum of the rotating shell and three other parameters, namely, $M_s$ the mass of the shell, $\sigma$ the surface density of the shell and $\omega$, the angular velocity of the rotating shell. Eq. (\ref{w1}) is a generalization of gravastar dynamical equation and reduces~\cite{Visser:2003ge,Usmani:2010ac} immediately to that if $\omega=0$ and $\Omega_{\pm}=0$.

Interesting extensions of the present analysis include the following:
\begin{itemize}
    \item The investigation of lightlike geodesics that enter the interior spacetime through the shell, searching for observational signatures of such configurations\cite{Chirenti:2007mk}.
    \item The consideration of the evolution of perturbed non-spherical shell perturbations that may also be connected with the emission of gravitational waves which may also provide observational signatures of such configurations.
    \item The investigation of physical mechanisms that may lead to the formation of such objects especially in the context of stabilized vacuum shell configurations forming during gravitational collapse \cite{Joshi:2011zm}.
    \item The consideration of stabilized thin shell configurations where the stability is induced by modified gravity, charge or other mechanisms \cite{Eiroa:2008hv,Horvat:2008ch,Lemos:2008aj,Richarte:2007zz}.
\end{itemize}

In conclusion, the new results presented in our analysis for the dynamics and stability of rotating shells provide new insights for such configurations which motivate the further research on the observational effects, properties and formation mechanisms for these objects.

\subsection{Acknowledgments}
This project was supported by the Hellenic Foundation for Research and Innovation (H.F.R.I.), under the "First call for H.F.R.I. Research Projects to support Faculty members and Researchers and the procurement of high-cost research
equipment Grant" (Project Number: 789).

\bibliographystyle{ws-ijmpd}
\bibliography{sample}

\end{document}